# Quality heterostructures from two dimensional crystals unstable in air by their assembly in inert atmosphere

Y. Cao[1,2], A. Mishchenko[1], G. L. Yu[1], K. Khestanova[2], A. Rooney[3], E. Prestat[3], A. V. Kretinin[1,2], P. Blake[4], M. B. Shalom[1,2], G. Balakrishnan[5], I. V. Grigorieva[1], K. S. Novoselov[1], B. A. Piot[6], M. Potemski[6], K. Watanabe[7], T. Taniguchi[7], S. J. Haigh[3], A. K. Geim[1,2], R. V. Gorbachev[1,2]

(1) School of Physics and Astronomy, University of Manchester, Oxford Road, Manchester M13 9PL, United Kingdom

(2) Manchester Centre for Mesoscience and Nanotechnology, University of Manchester, Manchester M13 9PL, United Kingdom

(3) School of Materials, University of Manchester, Manchester M13 9PL, United Kingdom

(4) Graphene Industries Ltd., 2 Tupelo Street, Manchester, M13 9HQ, UK

(5) Department of Physics, University of Warwick, Coventry CV4 7AL, United Kingdom

(6) Laboratoire National des Champs Magnétiques Intenses, CNRS-UJF-UPS-INSA, F-38042 Grenoble, France

(7) National Institute for Materials Science, 1-1 Namiki, Tsukuba, 305-0044 Japan

**Many layered materials can be cleaved down to individual atomic planes, similar to graphene, but only a small minority of them are stable under ambient conditions. The rest reacts and decomposes in air, which has severely hindered their investigation and possible uses. Here we introduce a remedial approach based on cleavage, transfer, alignment and encapsulation of air-sensitive crystals, all inside a controlled inert atmosphere. To illustrate the technology, we choose two archetypal two-dimensional crystals unstable in air: black phosphorus and niobium diselenide. Our field-effect devices made from their monolayers are conductive and fully stable under ambient conditions, in contrast to the counterparts processed in air. $NbSe_2$ remains superconducting down to the monolayer thickness. Starting with a trilayer, phosphorene devices reach sufficiently high mobilities to exhibit Landau quantization. The approach offers a venue to significantly expand the range of experimentally accessible two-dimensional crystals and their heterostructures.**

During recent years increasing attention has been paid to various two-dimensional (2D) crystals that can be mechanically exfoliated down to a monolayer or grown as monolayers on sacrificial substrates.[1-3] These materials offer a wide range of properties and include metals, semiconductors, normal and topological insulators, superconductors, etc. Moreover, individual 2D crystals can be combined in van der Waals (vdW) heterostructures,[1] which allows a way to expand much further the range of possible functionalities and assessable scientific problems. Unfortunately, current efforts in this direction are severely limited by the fact that many of 2D crystals rapidly degrade in air, reacting usually with oxygen and/or moisture.[4-8] As a result, the research field has so far been revolving mainly around a handful of highly stable and chemically inert monolayers such as graphene, hexagonal boron nitride (hBN) and several semiconducting dichalcogenides ($MoS_2$, $WSe_2$ and similar). To overcome the problem with poor stability of other interesting monolayers, we have developed a fabrication technique that allows air-sensitive crystals to be handled entirely under oxygen and moisture free conditions. In this report, we demonstrate capabilities of this approach using 2D crystals of black phosphorus (BP) and niobium diselenide.

Bulk NbSe$_2$ consists of covalently bonded Se–Nb–Se layers assembled in A-B-A-B fashion by weak (vdW-like) Se-Se bonding. It is a type-II s-wave superconductor with the transition temperature, $T_C$, of 7.2 K. There have been four previous studies of cleaved 2D niobium diselenide.[2, 9-11] The 1970s investigation[9] used changes in resistivity $\rho$ to estimate the number of layers, $N$, in the measured NbSe$_2$ crystallites and, as a result, underestimated their real thickness by one-two orders of magnitude. In other studies, NbSe$_2$ was found to be conductive down to a monolayer[2] and some traces of superconductivity were reported even in a 2-3 layer device.[10] In contrast, the most recent paper found that thin NbSe$_2$ deteriorated greatly, disallowing devices thinner than several nanometres.[11] The latter observation agrees with our own efforts to study superconductivity in 2D NbSe$_2$ using the standard microfabrication procedures that work well for graphene and air-stable dichalcogenides[1-3] (see below). The disagreement between different studies can be attributed to either accidental polymer contamination that coated and protected NbSe$_2$ in refs. [2, 10] or the use of lithographic techniques in ref. [11], which enhances oxidation during UV exposure.[12]

The second selected material, BP, has attracted huge interest during the last year.[4, 6-8, 13-21] Its atomic layers have a hexagonal structure, similar to graphene, but strongly corrugated.[13] The changes in symmetry make BP a semiconductor with a direct band gap that strongly depends on the number of layers.[5] Relatively thick crystals ($\approx$10 nm or 15 layers) have been shown to survive the standard lithographic procedures and show hole mobilities $\mu$ of up to 4,000 cm$^2$ V$^{-1}$ s$^{-1}$ (refs. [6,28]). For thinner devices, $\mu$ rapidly decreases with $N$ down to 10 – 100 cm$^2$ V$^{-1}$ s$^{-1}$ for few-layer BP.[15] Due to reactivity with oxygen and moisture, thin BP is found to rapidly develop a porous structure, which can occur in a matter of minutes, and eventually decomposes.[4, 5, 16] Mono- and bi- layer crystals of BP were studied by atomic force microscopy (AFM) and optical spectroscopy[17, 18] but their transport measurements are lacking. Moreover, the reports of poor stability have raised questions whether some of those results refer to intrinsic phosphorene or a partially decomposed material.[4, 5, 16]

**Glove box assembly**

To avoid oxygen and moisture, we have employed an argon environment provided by a glove box with levels of H$_2$O and O$_2$ below 0.1 ppm. The relatively small size of available 2D crystals necessitates micrometre-scale precision in their positioning during transfer and encapsulation. Such positioning is impossible using commercially available glove boxes, which permit high-resolution optical microscopy but always rely on manual operation. To this end, a fully motorized micromanipulation station has been specially built, which allows us to cleave, align and transfer micrometre-sized crystals robotically, using translation stages and micromanipulators, all programmed and operated remotely from the outside with joysticks.

As the first step during typical glove-box assembly, an air-sensitive bulk layered material is exfoliated onto a thin PMMA film. Then, a 2D crystal chosen in an optical microscope is lifted with monolayer hBN using a technique described in ref. [22] but with remote control. The resulting stack is then deposited typically onto a relatively large hBN crystal residing on a chosen substrate. The encapsulating hBN crystals are impermeable to all gases and liquids[23, 24] and provide permanent protection for air-sensitive 2D crystals against degradation. The final assembly can be taken out of the glove box to be either investigated or processed further into devices for transport measurements using common lithographic techniques.

Importantly, the monolayer hBN covers the assembled vdW heterostructures from the top and, for its relatively high tunnel conductivity,[25] allows electrical contacts by directly evaporating metal films onto it, without etching openings in the encapsulation. The achieved contact resistances using standard thin films (5nm Cr/ 100 nm Au) are between 1 and 10 k$\Omega$, depending on encapsulated material, with the highest values observed for semiconducting 2D crystals such as few-layer BP. Reactive ion etching can later be employed to define multiterminal mesas of a desired geometry. We did not notice degradation due to the resulting open edges over many weeks of storage and measurements.

**Experimental devices**

We have successfully studied electron transport in many BP and NbSe$_2$ devices down to a monolayer in thickness. Schematics of our devices are explained in Fig. 1. In the case of BP, we have used the standard assembly described above: 20-100 nm hBN as an atomically flat substrate[22, 26] and monolayer hBN as the encapsulating layer (Fig. 1a). It is easy to confuse been mono- and bi- layer hBN[31] and, occasionally, our devices had bilayer hBN as the top layer. This resulted in a contact resistance by a factor of 10 higher, which was acceptable although complicated the transport measurements. The whole multilayer stack was placed on top of an oxidized Si wafer (300 nm of SiO$_2$) which served as a back gate in our transport studies.

For 2D NbSe$_2$, we often opted for monolayer graphene as the encapsulating top layer (Fig. 1c) because it is much easier to visualize and transfer graphene than even thicker, few-layer hBN[27]. The parallel conductance through graphene encapsulation has little effect on superconductivity in NbSe$_2$, which has much higher carrier density per monolayer than graphene. To confirm that such influence is indeed minor, we have also fabricated two devices (3 and 4 layers of NbSe$_2$) encapsulated with monolayer hBN. The reference devices exhibited practically the same $\rho$ and $T_C$ as graphene-encapsulated NbSe$_2$ of the same thickness. For consistency, we refer below to the latter devices. As a substrate for NbSe$_2$ we used both hBN (on top of the oxidized Si wafer) and bare SrTiO$_3$ (STO), which again yielded similar results. The latter substrates were made from commercially available polished wafers (500 μm in thickness) which were annealed at 950°C in oxygen. The annealing resulted in large (~1 μm wide) atomically flat terraces.[28] Below we present mainly results for NbSe$_2$ on STO, because of considerable current interest in high-temperature superconductivity observed in conceptually similar structures (monolayer FeSe grown on STO) in which $T_C \approx 100$ K is often attributed to STO's high dielectric constant.[29]

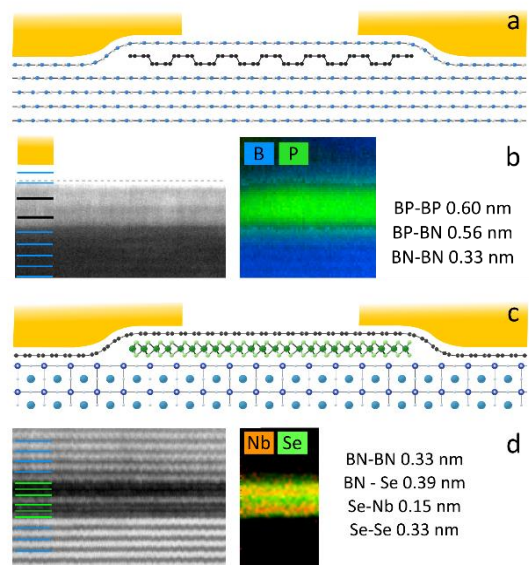

*Figure 1: Glove-box encapsulated air-sensitive 2D crystals.* (a) Schematics of the studied BP devices: mono- or few- layer BP is placed on an hBN substrate and protected with hBN. (b) Left – High angle annular dark-field STEM image of a heterostructure consisting of bulk hBN / bilayer BP / bilayer hBN and a top gold contact.

*Interlayer distances found from such images are stated on the right. Centre – STEM image superimposed with elemental profiles for the boron-K edge (blue) and phosphorus-$L_{2,3}$ edge (green) which are found using EELS for the same area. (c) Schematics of our $NbSe_2$ devices: 2D $NbSe_2$ is placed on hBN or atomically flat STO and covered with monolayer graphene. (d) Left – Cross-sectional bright field STEM image of a bulk hBN / bilayer $NbSe_2$/ bulk hBN with the extracted interlayer distances on the right. Centre – Superimposed STEM-EELS image with elemental profiles for the niobium–$M_{4,5}$ edge (orange) and selenium-$L_3$ edge (green).*

**Cross-sectional characterization**

It has previously been reported that contamination trapped between 2D crystals of graphene, hBN and stable dichalcogenides tends to segregate into isolated pockets leaving behind large areas with atomically clean and sharp interfaces.[30] The self-cleaning is however not a universal process and does not occur for all interfaces (for example, not for those involving 2D oxides).[22] In this work, we have observed self-cleaning of both BP and $NbSe_2$ surfaces in contact with graphene and hBN so that pockets of aggregated contamination clearly appear during glove-box assembly.[22, 30] Accordingly, our Hall bar devices were lithographically defined away from contamination pockets.[30] To confirm that the device interfaces are atomically clean, we have used a $Ga^+$ focused ion beam (FIB) to prepare thin (50 nm) cross-sections from some of our devices after their transport measurements. Figs. 1b,d show representative scanning transmission electron microscopy (STEM) images for 2D crystals of BP and $NbSe_2$, which are also accompanied by their cross-sections' analysis using electron energy loss spectroscopy (EELS).

The STEM image in Fig. 1d shows bilayer $NbSe_2$ sandwiched between two hBN crystals. Analysis of the interlayer separations yields values close to those in bulk $NbSe_2$ and hBN, which indicates little or no contamination trapped between the assembled layers. On the other hand, STEM studies of BP heterostructures were hindered by their high sensitivity to air. Although our cross-sectional specimens were exposed to air only for a few minutes during their transfer from FIB to STEM, BP heterostructures (unlike $NbSe_2$) already showed significant structural deterioration. In most of the studied specimens, bilayer BP was found to expand fourfold and lost any signs of a layered structure. Only in the contact regions where the heterostructures were covered with gold, we succeeded to find the BP layer nearly intact. In Fig. 1b the following layer sequence can be clearly seen: bulk hBN / bilayer BP / bilayer hBN / bulk Au. The interlayer distances calculated from the images for hBN are close to those in the bulk, within our experimental accuracy. However, the interlayer distance in BP is $\approx 0.6$ nm, somewhat larger than 0.55 nm for bulk BP. We attribute this enlargement to initial stages of degradation of our cross-sectional specimens.

It is known that the exact thickness of exfoliated 2D crystals is difficult to measure accurately using AFM, which yields significant variations in apparent thickness. This has been attributed to the presence of an additional contamination ('dead' layer) between 2D crystals and substrates and/or differences in their mechanical and adhesive properties with respect to the AFM tip (see, e.g., ref. [[2]]). For our devices with self-cleaned interfaces and covered with a protective monolayer over atomic steps, we have found their height to follow exactly the layer number *N* multiplied by the bulk interlayer distance with no evidence for any dead layer (Fig. 2a).

**Electron transport in encapsulated 2D black phosphorus**

We have studied more than 10 field-effect BP devices with thicknesses ranging from a monolayer to $N \approx 20$. All the devices exhibited ambipolar behaviour such that both hole and electron gases could be induced by applying sufficiently high gate voltage $V_g$. No degradation in the devices quality with time was noticed over the whole period of investigations lasting for several months. A summary of our transport studies for BP is given in Fig. 2. One can see that the response to field-effect doping is strongly asymmetric (Fig. 2b) with much higher conductivities σ and, hence, higher $\mu$ observed for holes, in agreement with the previous reports.[4, 6-8, 13-21] At liquid helium temperatures ($T$), field-effect mobilities measured in the 4-probe geometry were found to reach >4,000 cm$^2$ V$^{-1}$ s$^{-1}$ for our bulk-like devices (> 10 layers) but $\mu$ became progressively lower for thinner crystals, typically ≈1,200, 80 and 1 cm$^2$ V$^{-1}$ s$^{-1}$ for tri-, bi- and mono- layer devices, respectively (Fig. 2b). Note that gate-induced charge carriers in BP reside within a near-surface quantum well that has a width of only a few layers[6, 21] and, therefore, devices thicker than several atomic layers should conceptually be viewed as bulk BP with near-surface 2D gases.

Trilayer and thicker devices exhibited strong $T$ dependence of $\sigma$, which indicates that room-$T$ $\mu$ of BP is limited to ≤1,000 cm$^2$ V$^{-1}$ s$^{-1}$ because of phonon scattering. Low-$T$ $\mu$ found from Hall effect measurements were by a factor of ~ 2 lower than those determined from the electric field effect for all $N$, in agreement with previous reports.[4, 6-8, 13-21] Close to room $T$, differences in Hall and field effect $\mu$ diminish (inset in Fig. 2b). As for dominant scattering mechanism at low $T$, we can safely rule out charged impurities in hBN and, probably, contamination left at the hBN-BP interfaces. Most likely, the observed low-$T$ $\mu$ are limited by structural defects and strain induced during cleavage and transfer.

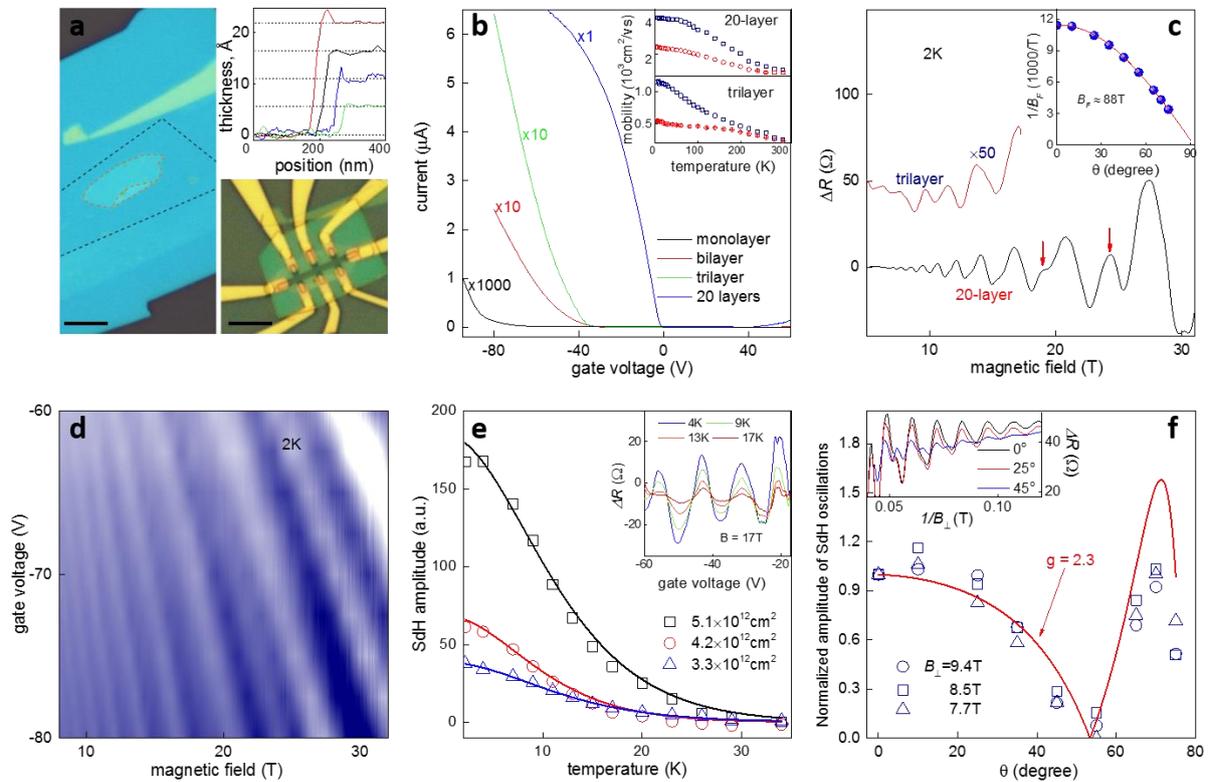

*Figure 2: Transport properties of atomically thin BP. (a) Left - Trilayer BP crystal (outlined in red and partially folded) is encapsulated with monolayer hBN outlined with the black lines. Right top – AFM measurements of thickness for several encapsulated BP samples. The dashed lines correspond to the interlayer spacing of 5.5 Å. Bottom – Optical micrograph of our typical Hall bar devices. Scale bars: 5 µm. (b) Source-drain current at 10 K as a function of $V_g$ for BP devices of different thickness, bias voltage 30 mV. Inset: T dependences of µ found using Hall and field effect measurements (red and blue curves, respectively) for 3- and 20- layer BP. (c) Changes in resistance ($\Delta R$) for the devices in (b). The red arrows mark spin-splitting of Landau levels. Inset: Angle dependence of the oscillation frequency. (d) Colour map $\rho(V_g,B)$ for the 20-layer device. Navy to white: $\rho$ changes by 115 Ohm. (e) T dependence of SdH oscillations. Inset: Examples at different temperatures. Their amplitude can be fitted by the Lifshitz–Kosevich formula (solid curves in the main plot). (f) Amplitude of SdH oscillations in the 20-layer BP for different orientations of magnetic field. Examples are shown in the inset.*

Two recent papers reported Shubnikov-de Haas (SdH) oscillations in BP devices thicker than 10 layers.[6, 20] The relatively high $\mu$ achieved for our glove-box encapsulated devices allowed the observation of SdH oscillations even for trilayer BP (Fig. 2c). The oscillations become visible in magnetic fields $B$ above a few Tesla, in agreement with the measured field-effect $\mu$. Their period, $1/B_F$, is found to match the carrier concentration $n$ estimated from the flat capacitor model, which shows that the majority of gate-induced carriers participate in electron transport. In order to confirm the two-dimensional nature of carriers in 2D BP, we have studied $B_F$ as a function of the angle $\theta$ between $B$ and the axis perpendicular to the BP plane. The data follow the expected $\cos(\theta)$ dependence that describes the field component perpendicular to the 2D plane, $B_\perp$ (inset of Fig. 2c). Furthermore, we have carried out analysis of $T$ dependence of SdH oscillations' amplitude and, using Lifshitz–Kosevich formula, determined the effective mass for holes as $m_h$ = 0.24 ±0.02 $m_0$ where $m_0$ is the free electron mass (Fig. 2e). This is in good agreement with *ab initio* values expected for few-layer BP[6] but notably smaller that $m_h \approx 0.3$ and 0.35 $m_0$ reported in refs. [6, 20], respectively. Note that all the reported values are smaller than $m_h \approx 0.42 m_0$ found for bulk BP.[6]

For devices with $N > 10$ layers we also observed lifting of the spin degeneracy in $B$ above 18 T (Fig. 2d). The Zeeman energy depends on the total magnetic field $B$ whereas the Landau level separation is determined by $B_\perp$. This allows one to distinguish between the two contributions and find the $g$-factor by studying angular dependence of SdH oscillations. Figure 2f shows their amplitude normalised by the value in the perpendicular field. Following the analysis suggested in ref. [31], we can fit the observed angular dependence using a single parameter $g$, which allows us to determine the $g$-factor value for the near-surface 2D hole gas as $\approx 2.3 \pm 0.2$.

Finally, to qualitatively illustrate how essential the glove-box encapsulation can be, we have tested photoluminescence (PL) from our 2D BP devices. No signal could be detected in any of them. To reconcile this observation with several recent reports on strong luminescence from few-layer BP samples, we have prepared partially encapsulated crystals such as the one shown in Fig. 3a. Strong luminescence could easily be detected from areas of thin BP crystals, which were exposed to air. The observed spectra (Fig. 3d) are also similar to those reported earlier for few-layer BP crystals (see, e.g., refs. [5, 17]). These observations suggest that the reported PL signals probably require some disorder induced by exposure to air. Furthermore, we have found that scanning with even a mild laser power of < 100 μW strongly increased the decomposition of BP, so that after only a few scans the entire non-encapsulated region was no longer visible in an optical microscope and ceased to yield any PL signal (Fig. 3d). This agrees well with the reports on optically-stimulated etching of BP in air.[5, 16] In contrast, encapsulated regions did not exhibit any changes even after many hours of continuous laser light exposure at ∼ 1 mW.

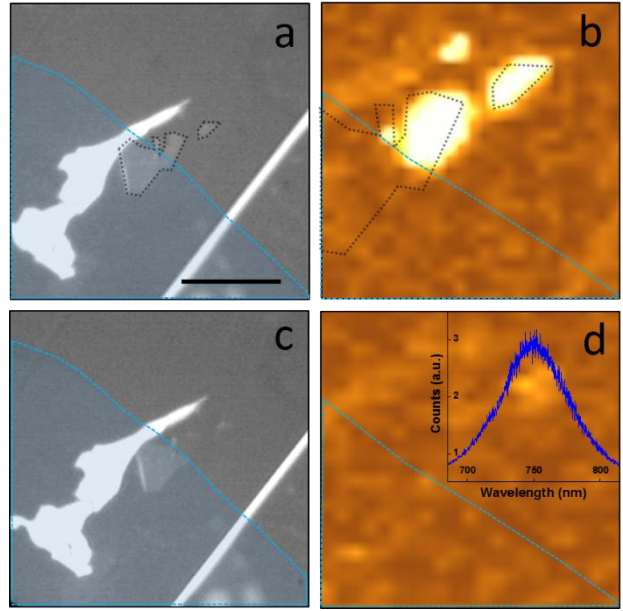

*Figure 3: Only non-protected few-layer BP shows strong photoluminescence.* (a) Bilayer crystal (outlined with black dots) is partially covered with few-layer hBN (semi-transparent blue area with the edge marked by the blue curve). Scale bar: 10 μm. (b) Photoluminescence map for this sample. No signal comes from the encapsulated BP but thin areas exposed to air are strongly luminescent. (c) After a few scans, clear structural changes occur in the non-protected part of the same BP crystal, which is attributed to photo-activated degradation.[5] (d) Subsequent scans of the damaged area yield no signal. Inset: Spectrum typical for few-layer BP during the first scan as in (b).

**Superconducting properties of few-layer NbSe$_2$**

To provide a reference for glove-box encapsulation, several NbSe$_2$ devices, all thinner than 10 layers, were made by the standard encapsulation with hBN but in air. All of them were found to be nonconductive, in agreement with the results reported in ref. [11]. In contrast, all the devices prepared using glove-box encapsulation (Fig. 4a) exhibited metallic behaviour of their ρ as a function of *T* (Fig. 4b) and a clear superconducting transition with a zero resistance (*R*) state (Figs. 4b, c). The transition is remarkably sharp (Fig. 4c), and the derivative *dR*/*dT* found by numerical differentiation of the measured resistance curves exhibits a single symmetric peak. The latter indicates high homogeneity of the devices (cf. ref. [11]). The critical temperature (defined experimentally as the peak position in *dR*/*dT*) gradually decreases with decreasing *N* (inset of Fig. 4c) but the suppression is relatively small even for the bilayer, indicating that 2D NbSe$_2$ remains largely free of disorder (cf. strong suppression of $T_c$ for few-layer Pb films[32]). The robust $T_c$ is consistent with the fact that the normal-state *R* of our devices (Fig. 4c) remains well below the threshold given by the resistance quantum, $h/4e^2$, above which superconductivity is expected to be strongly suppressed by disorder.[33-35] Only for monolayer NbSe$_2$ do we observe a significant drop in $T_c$ to values below 2 K, even though the 2D superconductor still remains very far from the resistance-quantum threshold. The origin of the observed rapid changes of $T_c$ in few-layer superconductors remains to be understood.

Furthermore, we have studied the influence of electric-field doping on $T_c$ for 2D NbSe$_2$ (Fig. 4c, d). To this end, we applied a gate voltage, $V_g$, to the metal film deposited on the back side of our 500 μm thick STO wafers. This is possible because STO has an extremely high dielectric constant (~10$^4$) at liquid helium temperatures.[36, 37] For our particular wafers, we have calibrated their dielectric response as a function of both *T* and $V_g$ by employing Hall bar devices made from graphene placed on top of the same STO wafers. The measurements yielded the field-effect density $\Delta n(V_g)$ shown in the upper x-axis of Fig. 4d. This axis shows that by applying $V_g > 50$ V at $T < 10$ K we induced $\Delta n \approx 1.3 \times 10^{13}$ cm$^{-2}$. Above this voltage, the dielectric constant of STO drops to such low values that a further increase in voltage results in little additional doping.[37] The field-induced $\Delta n$ should be compared with the total carrier density $n$ of intrinsic charge carriers in NbSe$_2$. Literature values for bulk NbSe$_2$ yield $n_L \approx 0.9 \times 10^{15}$ cm$^{-2}$ per monolayer. Our Hall effect measurements employing the two NbSe$_2$ devices encapsulated with hBN found $n = N \times 1.1 \pm 0.1 \times 10^{15}$ cm$^{-2}$, which indicates little change in carrier density for glove-box encapsulated 2D NbSe$_2$ with respect to bulk NbSe$_2$.

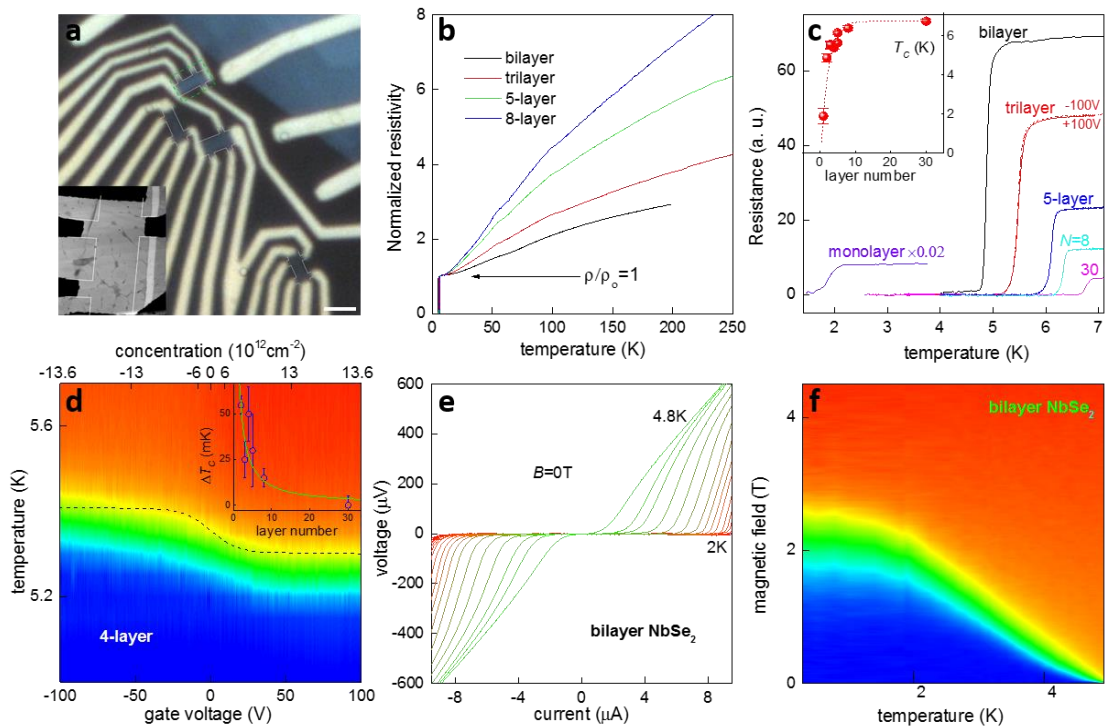

*Figure 4: Electron transport and superconductivity in few-layer NbSe$_2$.* (a) Optical micrograph showing 4 multiterminal NbSe$_2$ devices encapsulated with graphene. Scale bar: 2 μm. Inset: AFM topography for one of the devices (outlined with the green box in the main image) before its etching to define the Hall bar mesa (its later position is indicated by the white lines). (b) Resistance as a function of T for representative devices with different N. Resistivity is normalized to its value near the superconducting transition. (c) Changes in $T_C$ with N. The dotted line in the inset is a guide to the eye. (d) Example of maps $R(V_g,T)$ measured near the superconducting transition for N = 4. Blue to red: 0 to 20 Ohm. Yellow corresponds to the middle of the resistive transition. The black dashed curve shows its expected shift given by eq. (1). The upper x-axis is nonlinear and corresponds to the field-induced $\Delta n$. Inset: Changes in the critical temperature $\Delta T_C$ for different N, which correspond to swapping $V_g$ between 0 and +100 V. The green curve shows the expected functional dependence $\Delta T_C \propto 1/N$. (e) Critical current at different T for bilayer NbSe$_2$. (f) Superconducting phase diagram $R(B,T)$ for the same device. Blue to red: 0 to 80 Ohm. All the presented data are for NbSe$_2$ on STO and devices encapsulated with monolayer graphene (see section Experimental devices).

Our 2D NbSe$_2$ devices show small but notable changes in $T_C$ with applied gate voltage, despite the maximum electric-field doping reaches only ~1% of $n_L$. One can describe the observed changes by the Bardeen-Cooper-Schrieffer (BCS) theory that suggests[10]

$$\Delta T_C(V_g) = T_C(0) \times \frac{\Delta n}{n} \ln\left(\frac{T_C(0)}{1.14 T_D}\right), \tag{1}$$

where $T_D \approx 225$ K is the Debye temperature in bulk NbSe$_2$ and $T_C(0)$ the observed critical temperature for different $N$ at zero gate voltage. Eq. (1) implies that $T_C$ should vary linearly with induced $\Delta n$ and the expected behaviour of $T_C$ as a function of $V_g$ is shown by the dashed curve in Fig. 4d. The changes are strongly nonlinear and saturate at $V_g$ above ±50 V because of the STO dielectric response.[36, 37] One can see from Fig. 4d that the centre of the superconducting transition in $R(T)$, which is indicated by yellow, shifts with increasing $|V_g|$ and closely follows the dependence expected from eq. (1). Furthermore, according to (1), $\Delta T_C/T_C(0)$ should increase with decreasing the number of layers as $1/N$, and this functional dependence is shown by the solid green curve in the inset of Fig. 4d. The observed field-effect changes in $T_C$ qualitatively follow the theoretical dependences but their absolute values are approximately 3 times smaller than eq. (1) predicts. This is in contrast to the earlier measurements[10] where changes significantly larger than those expected in theory were reported for electric-field modulation of a partial superconducting transition in a NbSe$_2$ device with $N = 2 - 3$. The field-effect changes in $T_C$ reported in ref. [11] for thick NbSe$_2$ (~20 layer) strongly scattered but are in better agreement with our observations. Possible reasons for disagreement between experiments and theory have been discussed in ref. [10]. In addition, eq. (1) may need modification for this 2D limit, and the coupling constant and $T_D$ may change in few-layer NbSe$_2$ due to atomic thickness. Furthermore, let us point out that field-induced carriers reside within one monolayer closest to the gate. This means that any changes in superconductivity within this layer are affected by the proximity of other layers that do not experience any field effect. This observation agrees with the fact that the superconducting transition becomes progressively sharper with decreasing $N$ and, rather surprisingly, is sharpest for bilayer NbSe$_2$.

Finally, it is instructive to acknowledge recent reports where monolayer FeSe grown by molecular beam epitaxy on STO was found to exhibit $T_C$ reaching 100 K.[29, 32] The origin of this dramatic enhancement is not understood and, as mentioned above, the use of STO as a substrate in our experiments with mono- and few-layer NbSe$_2$ may provide some clues. For 2D NbSe$_2$ on STO, we observed only suppression of $T_C$ with respect to its bulk value and no difference with respect to using hBN as a substrate. The difference in electronic properties for FeSe and NbSe$_2$ is not expected to be crucial for any enhancement mechanism proposed so far. Nonetheless, it is also worth mentioning that we have carried out similar experiments using 4- and 6- layer FeSe that remain superconducting due to the glove-box encapsulation but FeSe on STO again showed reduced $T_C$, in qualitative agreement with the behaviour of 2D NbSe$_2$ of the same thickness. Therefore, the critical difference between our experiments and those reported in refs. [29, 32] could be different quality, disorder and strain. The use of non-doped STO in our experiments could also be important in considering possible enhancement mechanisms.

**Conclusion**

The described technology offers a universal platform for studying many new 2D crystals and vdW heterostructures, which so far could not be assessed experimentally because of their decomposition under ambient conditions and/or during lithography processing. The improvement provided by assembly in oxygen- and water- free environment is exemplified by our observation of superconductivity in monolayer NbSe$_2$ and the field-effect metallic conductivity for monolayer BP. They show no evidence of degradation and photoluminescence, even after keeping the encapsulated devices in air for many weeks. Despite the fact that the approach involves computer-controlled remote cleavage, positioning and handling of micron-sized monolayers, the technology is relatively straightforward, and we expect it to become wide spread within a short space of time.

**Methods**

**Preparation of thin cross sections** suitable for high-resolution STEM imaging was performed using an approach similar to that reported in ref. [30]. A dual-beam instrument (FEI Nova NanoLab 600), combining a field emission scanning electron microscope (SEM) and focussed ion beam (FIB) in the same chamber, has been used for site-specific preparation of cross-sectional samples using the lift-out approach.[38] Non-destructive SEM imaging of the device before milling allowed us to identify an area suitable for side-view imaging. Protective coatings of carbon (20 nm) and Au-Pd (30 nm) were sputtered onto the whole device surface *ex situ*, followed by an additional protective layer of Pt (2 µm thick) deposited *in situ* in the region of interest. Trenches were milled around the region of interest using a 30 kV Ga+ beam with a current of 1–10 nA to reduce thickness of the prospective slice to 1 µm. An Omniprobe™ micromanipulator was brought to touch the Pt protective top layer and secured there using additional Pt deposited *in situ* with the ion beam. The specimen lamella was then cut free from the substrate by ion milling and transferred to an Omniprobe™ TEM grid through a combination of Pt deposition and ion milling. The sample was thinned to near electron transparency with 30 kV $Ga^+$ at 0.1-0.5 nA. Final polishing steps at 5 kV (50 pA) and 2 kV (90 pA) reduced the lamella thickness to 20-70 nm and removed most of the amorphous surface layer that resulted from the higher energy milling. Site specific preparation was confirmed by comparing the region of interest viewed in SEM with the STEM image of the cross-sectional lamella, using gold contacts and pockets of contamination as features of reference.

**High resolution STEM imaging and spectroscopy** was performed using a probe side aberration-corrected FEI Titan G2 80-200 kV with an X-FEG electron source operated at 200 kV. Imaging and spectroscopy were carried out using a Gatan Imaging Filter (GIF) Quantum ER system with a 5 mm entrance aperture. High angle annular dark field (HAADF) and bright field STEM imaging was performed using a probe convergence angle of 21 mrad, a HAADF inner angle of 62 mrad and a probe current of ∼ 75 pA. Electron energy loss spectroscopy (EELS) was carried out using the same Gatan system with a 5 mm entrance aperture, a collection angle of 62 mrad and an energy dispersion of 1 eV. The multilayer structures were oriented along an <hkl0> crystallographic direction by taking advantage of the Kikuchi bands of the Si substrate.

**Photoluminescence (PL) spectroscopy** was performed at room temperature in air using a confocal Raman spectrometer (Witec). A laser spot of 0.5 µm in size was obtained using a Nikon 100x objective. Samples were illuminated with a 514.5 nm (2.41 eV) light at 100 µW intensity. Spatially resolved PL mapping was acquired using a piezoelectric stage, with an emission range of 690 to 810 nm (corresponding to a 600 g/mm spectrometer grating).

**References**

1. Geim AK, Grigorieva IV. Van der Waals heterostructures. *Nature* **499**, 419-425 (2013).

2. Novoselov KS, *et al.* Two-dimensional atomic crystals. *P. Natl. Acad. Sci. USA* **102**, 10451-10453 (2005).

3. Xu MS, Liang T, Shi MM, Chen HZ. Graphene-Like Two-Dimensional Materials. *Chem. Rev.* **113**, 3766-3798 (2013).

4. Koenig SP, Doganov RA, Schmidt H, Neto AHC, Ozyilmaz B. Electric field effect in ultrathin black phosphorus. *Appl. Phys. Lett.* **104**, 103106 (2014).

5. Castellanos-Gomez A, *et al.* Isolation and characterization of few-layer black phosphorus. *2D Materials* **1**, 025001 (2014).

6. Likai Li GJY, Vy Tran, Ruixiang Fei, Guorui Chen, Huichao Wang, Jian Wang, Kenji Watanabe, Takashi Taniguchi, Li Yang, Xian Hui Chen, Yuanbo Zhang. Quantum Oscillations in Black Phosphorus Two-dimensional Electron Gas. *arXiv:1411.6572*, (2014).

7. V. Tayari NH, I. Fakih, A. Favron, E. Gaufrès, G. Gervais, R. Martel, T. Szkopek. Two-Dimensional Magnetotransport in a Black Phosphorus Naked Quantum Well. *arXiv:1412.0259*, (2014).

8. Wood JD, *et al.* Effective Passivation of Exfoliated Black Phosphorus Transistors against Ambient Degradation. *Nano Lett.* **14**, 6964-6970 (2014).

9. Frindt RF. Superconductivity in Ultrathin $NbSe_2$ Layers. *Phys. Rev. Lett.* **28**, 299-301 (1972).

10. Staley NE, Wu J, Eklund P, Liu Y, Li LJ, Xu Z. Electric field effect on superconductivity in atomically thin flakes of $NbSe_2$. *Phys. Rev. B* **80**, 184505 (2009).

11. El-Bana MS, Wolverson D, Russo S, Balakrishnan G, Paul DM, Bending SJ. Superconductivity in two-dimensional $NbSe_2$ field effect transistors. *Supercond. Sci. Tech.* **26**, 125020 (2013).

12. Myers GE, Montet GL. Light-Induced Oxidation of $NbSe_2$ Single Crystals. *J. Phys. Chem. Solids* **32**, 2645-2646 (1971).

13. Brown A, Rundqvist S. Refinement of the crystal structure of black phosphorus. *Acta Crystallogr.* **19**, 684-685 (1965).

14. A. Favron EG, F. Fossard, P.L. Lévesque, A-L. Phaneuf-L'Heureux, N. Y-W. Tang, A. Loiseau, R. Leonelli, S. Francoeur, R. Martel. Exfoliating pristine black phosphorus down to the monolayer: photo-oxidation and electronic confinement effects. *arXiv:1408.0345*, (2014).

15. Li LK, *et al.* Black phosphorus field-effect transistors. *Nat. Nanotechnol.* **9**, 372-377 (2014).

16. Island JO, Steele GA, Zant HSJvd, Castellanos-Gomez A. Environmental instability of few-layer black phosphorus. *2D Materials* **2**, 011002 (2015).

17. Liu H, et al. Phosphorene: An Unexplored 2D Semiconductor with a High Hole Mobility. *Acs Nano* **8**, 4033-4041 (2014).

18. Xia FN, Wang H, Jia YC. Rediscovering black phosphorus as an anisotropic layered material for optoelectronics and electronics. *Nat. Commun.* **5**, 4458 (2014).

19. Joon-Seok Kim YL, Weinan Zhu, Seohee Kim, Di Wu, Li Tao, Ananth Dodabalapur, Keji Lai, Deji Akinwande. Toward Air-Stable Multilayer Phosphorene Thin-Films and Transistors. *arXiv:1412.0355*, (2014).

20. Gillgren N, et al. Gate tunable quantum oscillations in air-stable and high mobility few-layer phosphorene heterostructures. *2D Materials* **2**, 011001 (2015).

21. Low T, et al. Plasmons and Screening in Monolayer and Multilayer Black Phosphorus. *Phys. Rev. Lett.* **113**, 106802 (2014).

22. Kretinin AV, et al. Electronic Properties of Graphene Encapsulated with Different Two-Dimensional Atomic Crystals. *Nano Lett.* **14**, 3270-3276 (2014).

23. Bunch JS, et al. Impermeable atomic membranes from graphene sheets. *Nano Lett.* **8**, 2458-2462 (2008).

24. Hu S, et al. Proton transport through one-atom-thick crystals. *Nature* **516**, 227-230 (2014).

25. Britnell L, et al. Electron Tunneling through Ultrathin Boron Nitride Crystalline Barriers. *Nano Lett.* **12**, 1707-1710 (2012).

26. Dean CR, et al. Boron nitride substrates for high-quality graphene electronics. *Nat. Nanotechnol.* **5**, 722-726 (2010).

27. Gorbachev RV, et al. Hunting for Monolayer Boron Nitride: Optical and Raman Signatures. *Small* **7**, 465-468 (2011).

28. Koster G, Kropman BL, Rijnders GJHM, Blank DHA, Rogalla H. Quasi-ideal strontium titanate crystal surfaces through formation of strontium hydroxide. *Appl. Phys. Lett.* **73**, 2920-2922 (1998).

29. Ge J-F, et al. Superconductivity above 100 K in single-layer FeSe films on doped $SrTiO_3$. *Nat. Mater.* **Advance online publication**, (2014).

30. Haigh SJ, et al. Cross-sectional imaging of individual layers and buried interfaces of graphene-based heterostructures and superlattices. *Nat. Mater.* **11**, 764-767 (2012).

31. Knap W, *et al.* Spin and interaction effects in Shubnikov–de Haas oscillations and the quantum Hall effect in GaN/AlGaN heterostructures. *J. Phys. Condens. Matter* **16**, 3421-3432 (2004).

32. Qin SY, Kim J, Niu Q, Shih CK. Superconductivity at the Two-Dimensional Limit. *Science* **324**, 1314-1317 (2009).

33. Strongin M, Thompson RS, Kammerer OF, Crow JE. Destruction of Superconductivity in Disordered near-Monolayer Films. *Phys. Rev. B* **1**, 1078-1091 (1970).

34. Haviland DB, Liu Y, Goldman AM. Onset of Superconductivity in the Two-Dimensional Limit. *Phys. Rev. Lett.* **62**, 2180-2183 (1989).

35. Yazdani A, Kapitulnik A. Superconducting-Insulating Transition in Two-Dimensional a-MoGe Thin Films. *Phys. Rev. Lett.* **74**, 3037-3040 (1995).

36. Müller KA, Burkard H. $SrTiO_3$: An intrinsic quantum paraelectric below 4 K. *Phys. Rev. B* **19**, 3593-3602 (1979).

37. Dec J, Kleemann W, Westwanski B. Scaling behaviour of strontium titanate. *J. Phys. Condens. Matter* **11**, L379-L384 (1999).

38. Schaffer M, Schaffer B, Ramasse Q. Sample preparation for atomic-resolution STEM at low voltages by FIB. *Ultramicroscopy* **114**, 62-71 (2012).

**Acknowledgements** Bulk BP and $NbSe_2$ were purchased from HQ Graphene and 2D Semiconductors.